\documentstyle[11pt,preprint,amssymb,amsbsy,eqsecnum,aps,epsf]{revtex}

\renewcommand{\vec}{\boldsymbol}

\begin{document}
\everymath{\displaystyle}
\draft
\title{
Influence of guiding magnetic field on emission of stimulated photons in  generators utilizing periodic slow-wave structures.}
\author{V. G. Baryshevsky, K. G. Batrakov}
\address{
Institute of Nuclear Problems 220050, Minsk, Republic of Belarus}
\date{\today}
\maketitle

\begin{abstract}
Effect of guiding magnetic field on evolution of stimulated  emission is considered.  It is shown that the transverse dynamics of electrons  contributes to generation process and that contribution  decreases with the magnetic field grows. The  equation of generation for stimulated  radiation of electron beam passing over periodic medium in magnetic field of arbitrary value is derived. The critical value of guiding field for which the transverse dynamics of electron don't contribute to emission is determined. It is shown, that transverse dynamics of electron modifies the boundary conditions. It follows from the derived generation equation that  transverse dynamics yields to $\sim 25\% $  increase of the increment magnitude in high gain regime. In the limit of small signal the generation gain twice increases when transverse dynamics evolves. Obtained results are valid for every FEL system, which use the mechanism of waves slowing in the slow-wave structures for generation.   

\end{abstract}

\section{Introduction}
The  various types of devices utilizing the electron beam interaction with  electromagnetic fields in slow-wave system (Cherenkov \cite{Cher},  Smith-Purcell \cite{Smith}, quasi-Cherenkov \cite{Fer}, transition \cite{Trans}  radiation mechanisms) was considered in the past.The great number of researches in the area of microwave electronics  resulted in the traveling wave tube (TWT), backward-wave oscillator (BWO), orotrons and so on. Theory of such generators, as a rule, considers the one-dimentional (longitudinal) dynamics of electrons in the field of an electromagnetic wave. On the other hand in our previous works devoted to quasi-Cherenkov FEL  the stimulated emission of unmagnetized electron beam with three - dimensional dynamics was studied (\cite{1}, \cite{2}, \cite{3}). In our works \cite{4}, \cite{5} the guiding field was considered as strong and the electron beam dynamics as one-dimensional.
 It was shown that quasi-Cherenkov volume FEL (VFEL) can produce radiation with lower current density (especially in X-ray spectrum region of
frequencies) in comparison with ordinary FELs.
However, multiple scattering of electrons in the medium destructs
the coherence  of radiation process. The surface scheme of the
quasi-Cherenkov VFEL can be used for multiple scattering reducing (\cite{4}, 
\cite{5}). In that case an electron beam moves over a periodic medium at a
distance $d\leq \frac{u}{4\pi c}\gamma \lambda $ ($\lambda $ is the radiation wavelength, 
$\gamma =1/\sqrt{1-\frac{u^{2}}{c^{2}}}$ is the electron Lorentz factor) and
radiation is formed along the whole electron trajectory in vacuum without
multiple scattering. In these works (\cite{4}, \cite{5}) the generation of the
surface parametric FEL was considered for electron beam placed in strong
longitudinal guiding magnetic field. So, only the contribution of the
one-dimensional (longitudinal) dynamics of electrons to stimulated radiation was considered. This is true if inequality 
$\frac{eH^{(0)}}{m\gamma c\Delta }\gg 1$ is satisfied ($H^{(0)}$ is the magnitude of
magnetic field, $\Delta $ is the detuning from
synchronism condition). In opposite case $\frac{eH^{(0)}}{m\gamma c\Delta 
}\leq 1$ it is necessary to take into account the transverse motion of
electrons. The $\Delta $ increases with the interaction length $L_{\ast }$
decrease ( $L_{\ast }$ is the  length of interaction between the electron beam
and radiation) and with the current density increase. Therefore the transverse motion
contributes to stimulated quasi-Cherenkov emission in the case of high
current or small interaction length ($\frac{eH^{(0)}}{m\gamma c}\leq \max \{\frac{%
2\pi c}{L_{\ast }};(\omega \omega _{b}^{2}/\gamma ^{5})^{1/3}\}$, where $%
\omega _{b}=4\pi e^2 n_e /m_e$ is the Langmuir frequency).

In this paper we present the analysis of volume
FEL (VFEL) operation in the periodic slow-wave structures including the effect of the finite  guiding magnetic field value on stimulated emission. So we take into account the
contribution of transverse electron beam motion to generation process. Using the linearized
perturbation field approximation we derive  the boundary conditions,
dispersion relations and generation equation. All
results received below relate to Compton regime, when the amplitudes of
excited longitudinal Langmuir waves are small. Raman regime was   considered
in our previos work \cite{6} in two limiting cases 1)when guiding magnetic field is absent; 
2)when guiding field is strong.

\section{The problem statement}

\bigskip

\begin{figure}[htbp]
\epsfxsize=16 cm
\epsfbox{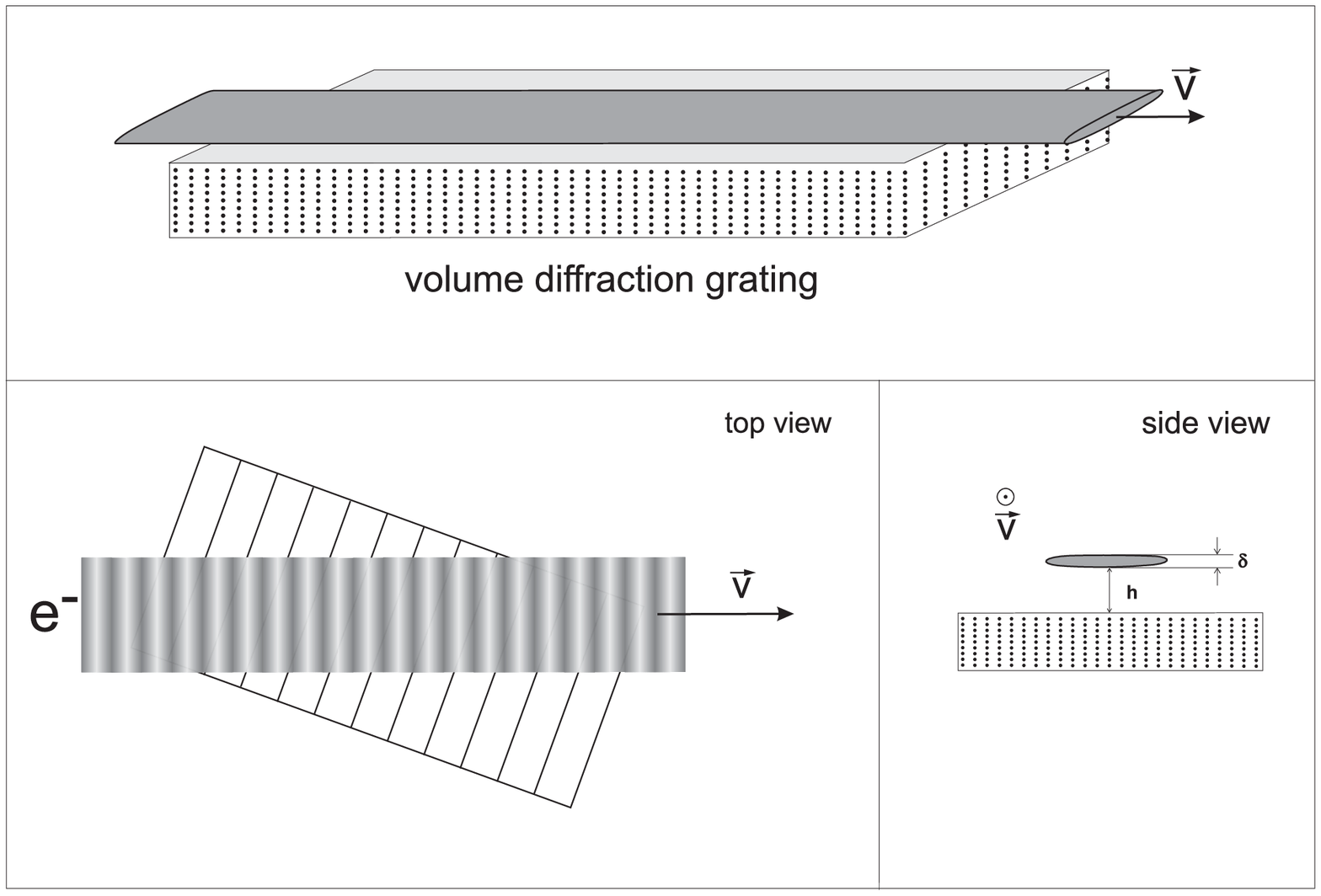}
\caption{Electron beam is moving over the periodic structure}
\label{fig.1}
\end{figure}

The considered system is represented on Figure \ref{fig.1}. An electron beam located
at height $h$ over the spatial periodic structure is moving
parallel to structure surface. $\delta $ is the beam thickness in the
direction normal to the surface. The axis $x$ is chosen along the
direction of electron motion, the axis $z$ is normal to the periodic
structure surface. Dynamical diffraction on the periodic structure forms the
3-dimensional volume distribution feedback which provides generation regime.
The set of reciprocal lattice vectors $\tau _n=\{ 2\pi n_1/d_1;2\pi n_2/d_2;2\pi n_3/d_3\}$ defining diffraction process can
be directed at an arbitrary angle relative to the particle velocity and to
the surface, ($d_i$ are translation periods of periodic structure, $n_i$ are intergers). The interaction between the electron beam and the grating
produces an emission spectrum. The emission frequency is defined by the spatial
period and by the volume geometry (by the direction of the reciprocal
vectors for example). The periodic structure perfoms two basic functions. Firstly it
slows down the phase velocity of an electromagnetic wave that enables 
conditions for coherent radiation. Secondly, due to 3-dimensional
distributed feedback, the periodic structure is an effective volume resonator,which gives the possibility for an oscillator regime realization. The
 emitted photons bunch the electron beam. This bunching
leads to greater emission, which leads to more bunching.Three-dimensional Bragg  distributed feedback
keeps emission in the interaction region. The regime of an oscillator is
realized as the result of these processes. Dependence of the emitted wavelength on
system geometry provides smooth frequency tuning. For deriving the
generation equation it is necessary to obtain the dispersion equations in
all regions and to use the boundary conditions on the surfaces of the
electron beam and surfaces of the slow-wave structure.

\section{Dispersion equations}

\bigskip

In most of previous works concerning slow-wave FELs the electron beam is considered
as magnetized. Therefore only longitudinal dynamics of electron beam
was taken into account. The magnetic field is used for electron beam guiding 
 over slow-wave structure surface. However, the transverse motion of electron still can
contribute to the process of stimulated radiation. The contribution of
transverse degrees of freedom depends on: 1) parameters of an electron beam
such as the energy of electrons, current density, the velocity
spread of an electron beam; 2) the parameters of emitted radiation such as
photon wavelength and the field amplitudes; 3) the parameters of the
electrodynamical structure such as photoabsoption length, interaction length of electron beam with emitted radiation and the binding of an electron beam with eigenmodes of an
electrodynamic system; 4) the magnitude of guiding field.

The slow electromagnetic wave which is in 
synhronism with the electron beam produces modulation of the density and
current density. This leads to development of  instability.
The stimulated  radiation is the result of this instability.
Let us consider the influence of guiding magnetic field on the stimulated
 radiation. The velocity and radius-vector of an electron
can be presented as: ${\mathbf{v}}_{\alpha }(t)={\mathbf{u}}+\delta {\mathbf{v}}_{\alpha
}(t)$, ${\mathbf{r}}_{\alpha }(t)={\mathbf{r}}_{0\alpha }+{\mathbf{u}}t+\delta {\mathbf{r}}%
_{\alpha }(t)$. Here the pertubations $\delta {\mathbf{v}}_{\alpha }(t)$ and $%
\delta r_{\alpha }(t)$ are results of electron interaction with an
electromagnetic wave.
In the linear field approximation the current density can be written as

\begin{eqnarray}
{\mathbf{j}}(z,k_{x},k_{y},\omega ) &=&e\sum\limits_{\alpha }\{{\mathbf{u}}[\delta
(z-z_{0\alpha })\exp (-ik_{x}x_{0\alpha }-ik_{y}y_{0\alpha })(-ik_{x}\delta
x_{\alpha }(\omega -k_{x}u)-  \nonumber \\
&&ik_{y}\delta y_{\alpha }(\omega -k_{x}u))-  \label{cur1} \\
&&\frac{\partial }{\partial z}\delta (z-z_{0\alpha })\delta z_{\alpha
}(\omega -k_{x}u)\exp (-ik_{x}x_{0\alpha }-ik_{y}y_{0\alpha })]+  \nonumber \\
&&\delta {\mathbf{v}}_{\alpha }(\omega -k_{x}u)\delta (z-z)\exp
(-ik_{x}x_{0\alpha }-ik_{y}y_{0\alpha })\},  \nonumber
\end{eqnarray}
where $x_{\alpha }(t)=x_{0\alpha }+ut+\delta x_{\alpha }(t)$, $y_{\alpha
}(t)=y_{0\alpha }+ut+\delta y_{\alpha }(t)$, $z_{\alpha }(t)=z_{0\alpha
}+\delta z_{\alpha }(t)$ are the radius vectors of an electrons in a beam
and $\{\delta x_{\alpha },\delta z_{\alpha },\delta {\mathbf{v}}_{\alpha
}\}(\omega )=\int dt\exp (i\omega t)\{\delta x_{\alpha },\delta y_{\alpha
},\delta z_{\alpha },\delta {\mathbf{v}}_{\alpha }\}(t)$, $\ {\mathbf{j}}%
(z,k_{x},k_{y},\omega )=\int dxdy\exp (-ik_{x}x-ik_{y}y){\mathbf{j}}(z,x,y,\omega
)$. Dynamics of electron in the field of electromagnetic wave is described
by equation

\begin{eqnarray}
\frac{d\delta {\mathbf{v}}_{\alpha }(t)}{dt}-\frac{e}{m\gamma c}[\delta {\mathbf{v}}
_{\alpha }(t){\mathbf{H}}_{0}] &=&\ \ \frac{e}{m\gamma }\{{\mathbf{E}}({\mathbf{r}}
_{\alpha }(t),t)+\frac{1}{c}[{\mathbf{u}}{\mathbf{H}}({\mathbf{r}}_{\alpha }(t),t)]-
\label{mot1} \\
\frac{{\mathbf{u}}}{c^{2}}({\mathbf{u}}{\mathbf{E}}({\mathbf{r}}_{\alpha }(t),t))\}.  \nonumber
\end{eqnarray}
The distinction from the schemes studied earlier (\cite{4},\cite{5}) is in considering the term
with guiding magnetic field ${\mathbf{H}}_{0}$. The Fourier transformation of
(\ref{mot1}) gives

\begin{eqnarray}
\delta {\mathbf{v}}_{\alpha }(\omega )-\frac{e}{m\gamma c}[\delta {\mathbf{v}}
_{\alpha }(\omega ){\mathbf{H}}_{0}]=\frac{ie}{m\gamma \omega }\int \frac{
dk_{x}^{\prime }dk_{y}^{\prime }}{(2\pi )^{2}}\exp (ik_{x}^{\prime
}x_{0\alpha }+ik_{y}^{\prime }y_{0\alpha }) \nonumber  \\
\{{\mathbf{E}}(z_{0\alpha },k_{x}^{\prime },k_{y}^{\prime },\omega -k_{x}^{\prime
}u)+\frac{1}{c}[{\mathbf{u}}{\mathbf{H}}(z_{0\alpha },k_{x}^{\prime },k_{y}^{\prime
},\omega +k_{x}^{\prime }u)]- \label{four1}  \\
\frac{{\mathbf{u}}}{c^{2}}({\mathbf{u}}{\mathbf{E}}(z_{0\alpha },k_{x}^{\prime
},k_{y}^{\prime },\omega +k_{x}^{\prime }u))\};\text{ \ \ }\delta {\mathbf{r}}%
_{\alpha }(\omega )=\frac{i}{\omega }\delta {\mathbf{v}}_{\alpha }(\omega ) 
\nonumber
\end{eqnarray}

Decomposing (\ref{four1}) by components it can be received

$
\begin{array}{l}
\delta v_{x\alpha }(\omega )=\frac{ie}{m\gamma ^{3}\omega }\int \frac{%
dk_{x}^{\prime }dk_{y}^{\prime }}{(2\pi )^{2}}\exp (ik_{x}^{\prime
}x_{0\alpha }+ik_{y}^{\prime }y_{0\alpha })E_{x}(z_{0\alpha },k_{x}^{\prime
},k_{y}^{\prime },\omega +k_{x}^{\prime }u) \\ 
\delta v_{y\alpha }(\omega )=\frac{1}{\omega ^{2}-\left( \frac{eH^{(0)}}{%
m\gamma c}\right) ^{2}}\int \frac{dk_{x}^{\prime }dk_{y}^{\prime }}{(2\pi
)^{2}}\exp (ik_{x}^{\prime }x_{0\alpha }+ik_{y}^{\prime }y_{0\alpha }) \\ 
\left\{ -\frac{e^{2}H^{(0)}}{m^{2}\gamma ^{2}c}\left\{ \frac{\omega }{%
\omega +k_{x}^{\prime }u}E_{z}-\frac{iu}{\omega +k_{x}^{\prime }u}\frac{%
\partial E_{x}}{\partial z}\right\} +\frac{ie\omega }{m\gamma }\left( 
\frac{\omega }{\omega +k_{x}^{\prime }u}E_{y}+\frac{k_{y}^{\prime }u}{%
\omega +k_{x}^{\prime }u}E_{x}\right) \right\} \\ 
\delta v_{z\alpha }(\omega )=\frac{1}{\omega ^{2}-\left( \frac{eH^{(0)}}{%
m\gamma c}\right) ^{2}}\int \frac{dk_{x}^{\prime }dk_{y}^{\prime }}{(2\pi
)^{2}}\exp (ik_{x}^{\prime }x_{0\alpha }+ik_{y}^{\prime }y_{0\alpha }) \\ 
\left\{ \frac{ie}{m\gamma }\omega \left\{ \frac{\omega }{\omega
+k_{x}^{\prime }u}E_{z}-\frac{iu}{\omega +k_{x}^{\prime }u}\frac{\partial
E_{x}}{\partial z}\right\} +\frac{e^{2}H^{(0)}}{m^{2}\gamma ^{2}c}\left( 
\frac{\omega }{\omega +k_{x}^{\prime }u}E_{y}+\frac{k_{y}^{\prime }u}{%
\omega +k_{x}^{\prime }u}E_{x}\right) \right\}
\end{array}
$

If electrons in the beam are distributed as $n=n_{e}f(z_{\alpha })$ it can
be derived from (\ref{cur1},\ref{four1})

\begin{eqnarray}
\delta {\mathbf{j}}(z,k_{x},k_{y},\omega )=\frac{\omega _{L}^{2}}{4\pi }%
f(z)\left\{ \frac{i\omega }{\gamma ^{3}(\omega -{\mathbf{k}}{\mathbf{u}})^{2}}E_{x}%
{\mathbf{e}}_{x}+\right.  \nonumber  \\
\frac{-\frac{eH^{(0)}}{m\gamma ^{2}c}\left\{ \frac{\omega -k_{x}u}{\omega 
}E_{z}-\frac{iu}{\omega }\frac{\partial E_{x}}{\partial z}\right\} +\frac{%
i(\omega -{\mathbf{k}}{\mathbf{u}})}{\gamma }\left( \frac{\omega -k_{x}u}{\omega }%
E_{y}+\frac{k_{y}u}{\omega }E_{x}\right) }{(\omega -{\mathbf{k}}{\mathbf{u}}%
)^{2}-\left( \frac{eH^{(0)}}{m\gamma c}\right) ^{2}}{\mathbf{e}}_{y}+  \nonumber \\
\left. \frac{\frac{i}{\gamma }(\omega -{\mathbf{k}}{\mathbf{u}})\left\{ \frac{%
\omega -k_{x}u}{\omega }E_{z}-\frac{iu}{\omega }\frac{\partial E_{x}}{%
\partial z}\right\} +\frac{eH^{(0)}}{m\gamma ^{2}c}\left( \frac{\omega
-k_{x}u}{\omega }E_{y}+\frac{k_{y}u}{\omega }E_{x}\right) }{(\omega -{\mathbf{k}}%
{\mathbf{u}})^{2}-\left( \frac{eH^{(0)}}{m\gamma c}\right) ^{2}}{\mathbf{e}}%
_{z}\right\} + \label{bas1}  \\
\frac{{\mathbf{u}}k_{y}}{\omega -{\mathbf{k}}{\mathbf{u}}}\frac{\omega _{L}^{2}}{4\pi }%
f(z)\frac{-\frac{eH^{(0)}}{m\gamma ^{2}c}\left\{ \frac{\omega -k_{x}u}{%
\omega }E_{z}-\frac{iu}{\omega }\frac{\partial E_{x}}{\partial z}\right\} +%
\frac{i(\omega -{\mathbf{k}}{\mathbf{u}})}{\gamma }\left( \frac{\omega -k_{x}u}{%
\omega }E_{y}+\frac{k_{y}u}{\omega }E_{x}\right) }{(\omega -{\mathbf{k}}{\mathbf{u}}%
)^{2}-\left( \frac{eH^{(0)}}{m\gamma c}\right) ^{2}}-  \nonumber \\
\frac{i\vec{u}}{\omega -{\mathbf{k}}{\mathbf{u}}}\frac{\omega _{L}^{2}}{4\pi }\frac{%
\partial }{\partial z}\left( f(z)\frac{\frac{i}{\gamma }(\omega -{\mathbf{k}}%
{\mathbf{u}})\left\{ \frac{\omega -k_{x}u}{\omega }E_{z}-\frac{iu}{\omega }%
\frac{\partial E_{x}}{\partial z}\right\} +\frac{eH^{(0)}}{m\gamma ^{2}c}%
\left( \frac{\omega -k_{x}u}{\omega }E_{y}+\frac{k_{y}u}{\omega }%
E_{x}\right) }{(\omega -{\mathbf{k}}{\mathbf{u}})^{2}-\left( \frac{eH^{(0)}}{m\gamma c%
}\right) ^{2}}\right)  \nonumber
\end{eqnarray}
Current density contains terms with Cherenkov and cyclotron resonances.
We shall study the Compton regime of Cherenkov instability. The terms corresponding to second order resonances give maximal contributions in that case. Below we use this fact for separation of wave polarisations.

The dispersion equation in the region filled with electron beam is defined
by equating of the determinant of the system  to zero.
\begin{equation}
(k^{2}c^{2}-\omega ^{2}){\mathbf{E}}-c^{2}{\mathbf{k}}({\mathbf{k}}{\mathbf{E}})=-4\pi i\omega
\delta {\mathbf{j}}({\mathbf{k}},\omega )\text{,}  \label{disp1}
\end{equation}
 Here $\delta {\mathbf{j}}({\mathbf{k}},\omega )$ is derived from (\ref{bas1})
($\delta {\mathbf{j}}(z,k_{x},k_{y},\omega )\sim \delta {\mathbf{j}}({\mathbf{k}},\omega
)\exp (ik_{z}z)$. It is considered in this case that $f(z)=1$ in the region
with electron beam).

Let us discuss some features of this dispersion equation. In general case of an
arbitrary guiding magnetic field it has six roots $k_{za}(k_{x},k_{y},\omega
)$, $a=1\div 6$. For the case of strong guiding magnetic field when the
condition $\omega -{\mathbf{k}}{\mathbf{u}}<<\frac{eH^{(0)}}{m\gamma c}$ there exist
four roots

\begin{eqnarray}
k_{bz} &=&\pm \sqrt{\frac{\omega ^{2}}{c^{2}}-k_{||}^{2}}\text{ when the
wave polarisation is normal to }{\mathbf{u}}\text{ and }{\mathbf{k}}  \label{sol1} \\
k_{bz} &=&\pm \sqrt{\left( \frac{\omega ^{2}}{c^{2}}-k_{||}^{2}\right)
\left( 1-\frac{\omega _{L}^{2}}{\gamma ^{3}(\omega -k_{x}u)^{2}}\right) }%
\text{ when the wave polarisation }  \nonumber \\
&&\text{is in the plane of }{\mathbf{u}}\text{ and }{\mathbf{k}}  \nonumber
\end{eqnarray}

The fist two roots correspond to electromagnetic waves which don't interact
with the electron beam. The last two roots correspond to waves which are
result of electromagnetic wave with electron beam interactions.In particular
case when $k_{z}=0$, these two wave degenerate to longitudinal slow and fast
Langmuir waves with the dispersion equation $1-\frac{\omega _{L}^{2}}{%
\gamma ^{3}(\omega -k_{x}u)^{2}}=0$.

In the opposite case of low guiding field, when inequality $\omega -{\mathbf{k}}%
{\mathbf{u}}>>\frac{eH^{(0)}}{m\gamma c}$ is satisfied there exist four roots of
(\ref{disp1})

\begin{eqnarray*}
k_{bz} &=&\pm \sqrt{\frac{\omega ^{2}}{c^{2}}-k_{||}^{2}-\frac{\omega
_{L}^{2}}{\gamma }}\text{ when the wave polarisation is normal to }{\mathbf{u}}%
\text{ and }{\mathbf {k}} \\
k_{bz} &=&\pm \sqrt{\frac{\omega ^{2}}{c^{2}}-k_{||}^{2}-\frac{\omega
_{L}^{2}}{\gamma }}\text{ \ when the wave polarisation is in the plane of }%
{\mathbf{u}}\text{ and }{\mathbf{k}}
\end{eqnarray*}
and the Langmuir waves polarized parallel to wavevector ${\mathbf{k}}$. So in the
region filled by beam ($h<z<h+\delta $) the field can be written as

\begin{equation}
\sum_{\{in\}}\left[ {\mathbf{e}}_{bn}^{(i)}a_{b}^{(n)}\exp (-ik_{bnz}z)\exp \{i(%
{\mathbf{k}}_{||}+{\mathbf{\tau}}_{n||}){\mathbf{r}}_{||}\}+{\mathbf{e}}_{bn}^{(i)}b_{b}^{(n)}%
\exp (ik_{bnz}z)\exp \{i({\mathbf{k}}_{||}+{\mathbf{\tau}}_{n||}){\mathbf{r}}_{||}\}\right]
\label{bb1}
\end{equation}

In vacuum regions 1 ($z>h+\delta $), 3 ($0<z<h$) and 5 ($z<-D$) the
electromagnetic field is a set of transverse polarized plane waves
\begin{equation}
\begin{array}{l}
\sum_{\{in\}}{\mathbf{e}}_{1n}^{(i)}a_{1}^{(n)}\exp (-ik_{nz}z)\exp \{i({\mathbf{k}}%
_{||}+{\mathbf{\tau}}_{n||}){\mathbf{r}}_{||}\}\text{ in region }1\text{ which is over
the electron beam}   \\
\sum_{\{in\}}\left[ {\mathbf{e}}_{3n}^{(i)}a_{3}^{(n)}\exp (-ik_{nz}z)\exp \{i(%
{\mathbf{k}}_{||}+{\mathbf{\tau}}_{n||}){\mathbf{r}}_{||}\}+{\mathbf{e}}_{3n}^{(i)}b_{3}^{(n)}%
\exp (ik_{nz}z)\exp \{i({\mathbf{k}}_{||}+{\mathbf{\tau}}_{n||}){\mathbf{r}}_{||}\}\right]\\ 
\text{ in the gap between the electron beam and slow-wave structure}   \\
\sum_{\{in\}}{\mathbf{e}}_{5n}^{(i)}a_{5}^{(n)}\exp (ik_{nz}z)\exp \{i({\mathbf{k}}%
_{||}+{\mathbf{\tau}}_{n||}){\mathbf{r}}_{||}\}\text{ in region }5  \\
\text{ which is under
the slow-wave structure}  
\end{array}
\label{vac1}
\end{equation}
Writing fields in region 1 and 5 we use the lack of the incident waves. The
electromagnetic field in the slow wave structure ($-D<z<0$) can be written
as a sum of Bloch functions

\begin{equation}
\sum_{\alpha }f_{\alpha }{\mathbf{E}}_{\alpha }({\mathbf{r}})=f_{\alpha }\exp \{i\vec{k%
}^{(\alpha )}{\mathbf{r}}\}{\mathbf{u}}_{\alpha }({\mathbf{r}}),  \label{str1}
\end{equation}
where ${\mathbf{u}}_{\alpha }({\mathbf{r}})$ satisfies to conditions 
${\mathbf{u}}_{\alpha}({\mathbf{r}}+{\mathbf{d}}_{i})={\mathbf{u}}_{\alpha }({\mathbf{r}})$ and ${\mathbf{d}}_{i}$  is arbitrary translation vector of spatially periodic slow-wave structure.

\section{The boundary conditions}

To derive the generation conditions  it is necessary 
to write the equations for field coefficient in (\ref{bb1},\ref{vac1},\ref
{str1}). These equations are produced by utilizing the boundary conditions
on the surfaces. If the surface currents and surface charges are not excited
on the boundary, then we shall use the conditions of transverse
magnetic and electric field continuity on the boundary.In general case, as will be shown below, the
induced surface currents and charges exist at the electron beam surfaces.
For defining of this currents and deriving of corresponding boundary
conditions the consideration of self-consistent problem of electron
beam-radiation interaction should be performed. To produce the boundary
condition for tangential component of magnetic field we use (\ref{bas1}) and 
Maxwell equation

\begin{equation}
rot\vec{H}=\frac{4\pi }{c}\delta {\mathbf{j}}  \label{maxw}
\end{equation}
By integrating left and right hand sides (\ref{maxw}) in narrow region
near the electron beam surface and using (\ref{bas1}), it can be derived the
following boundary conditions

\begin{eqnarray}
\left[ H_{y}+\frac{u}{c}\frac{\omega _{L}^{2}}{\gamma }f(z)\frac{\Delta
\left\{ \frac{\Delta }{\omega }E_{z}-\frac{iu}{\omega }\frac{\partial
E_{x}}{\partial z}\right\} -ia_{0}\left( \frac{\Delta }{\omega }E_{y}+%
\frac{k_{y}u}{\omega }E_{x}\right) }{\Delta D_{0}}\right] _{z_{b}}=0
\label{bound1} \\
\lbrack H_{x}]=0  \nonumber
\end{eqnarray}
Here the new symbols are introduced $\Delta =\omega -{\mathbf{k}}{\mathbf{u}}$, $%
D_{0}=(\omega -{\mathbf{k}}{\mathbf{u}})^{2}-\left( \frac{eH^{(0)}}{m\gamma c}\right)
^{2}$, $a_{0}=\frac{eH^{(0)}}{m\gamma c}$. As can be seen from (\ref{bound1}%
) the tangential component of magnetic field which is normal to electron
beam velocity ${\mathbf{u}}$ don't conserve on the electron beam surfaces.
The component of magnetic field parallel \ to velocity is conserved. The
nonconserving of $H_{y}$ on the electron beam density discontinuity is
caused by \ arising of surface current directed along the electron velocity
vector ${\mathbf{u}}$. The following limit cases of boundary conditions (\ref{bound1})
exist.

1) the limit of strong guiding magnetic field $\Delta <<a_{0}$, the $y$
component of magnetic field is conserved. That is result of transverse
dynamics lack in strong longitudinal magnetic field;

2) the opposite limit of weak guiding field $\Delta >>a_{0}$, in this case 
the boundary condition has the form:

\begin{equation}
\left[ H_{y}+\frac{u}{c}\frac{\omega _{L}^{2}}{\gamma }f(z)\frac{\frac{%
\Delta }{\omega }E_{z}-\frac{iu}{\omega }\frac{\partial E_{x}}{\partial z}%
}{\Delta ^{2}}\right] _{z_{b}}=0,  \label{bound2}
\end{equation}
and electron beam gives the resonant contribution to boundary
condition (\ref{bound2}) in the region of Cherenkov synhronism.

\section{The generation equations}

The scheme of a surface VFEL is shown in Fig 1. There $h$ is the distance
between an electron beam and a target surface, $\delta $ is a transverse
size of an electron beam.

The electromagnetic field excited in this system has the following form:

1) $z>h+\delta $

\begin{equation}
t\exp \{-ik_{z}(h+\delta )\}\exp \{i\vec{k}{\mathbf{r}}\}+\sum_{i}m_{i}\exp \{i%
{\mathbf{k}}_{i}{\mathbf{r}}\}
\end{equation}

2) $h<z<h+\delta $

\begin{eqnarray}
&&a\exp \{i\vec{k}_{b}{\mathbf{r}}\}+b\exp \{i\vec{k}_{b}^{(-)}{\mathbf{r}}\}+
\label{f2} \\
&&\sum_{i}m_{i}\exp \{i\vec{k}_{i}{\mathbf{r}}\}  \nonumber
\end{eqnarray}

3) $0<z<h$

\begin{equation}
c\exp \{i\vec{k}{\mathbf{r}}\}+d\exp \{-i\vec{k}^{(-)}{\mathbf{r}}\}+\sum_{i}m_{i}\exp
\{i\vec{k}_{i}{\mathbf{r}}\}  \label{f3}
\end{equation}

4) $-D<z<0$

\begin{equation}
\sum_{\alpha }f_{\alpha }\exp \{i\vec{k}^{(\alpha )}{\mathbf{r}}\}u_{\alpha }(%
{\mathbf{r}})  \label{f4}
\end{equation}

5) $z<-D$

\begin{equation}
\sum_{i}g_{i}\exp \{i\vec{k}_{i}^{(-)}{\mathbf{r}}\}  \label{f5}
\end{equation}
where ${\mathbf{k}}=({\mathbf{k}}_{\perp };k_{z})$, ${\mathbf{k}}^{(-)}=({\mathbf{k}}_{\perp
};-k_{z})$; $k_{z}=\sqrt{\omega ^{2}/c^{2}-k_{\perp }^{2}}$; ${\mathbf{k}}_{i}=(%
{\mathbf{k}}_{\perp }+{\mathbf{\tau}}_{i\perp },k_{iz})$, ${\mathbf{k}}_{i}^{(-)}=({\mathbf{k}}%
_{\perp }+{\mathbf{\tau}}_{i\perp },-k_{iz})$, $k_{iz}=\sqrt{\omega ^{2}/c^{2}-(%
{\mathbf{k}}_{\perp }+{\mathbf{\tau}}_{i\perp })^{2}}$, the wave vectors \{${\mathbf{k}}%
_{i} $\} and \{${\mathbf{k}}_{i}^{(-)}$\} correspond to electromagnetic waves
escaping from the system (if $k_{iz}$ is real) and evanescent waves (if $%
k_{iz}$ is imaginary).\{$F_{\alpha }=\exp \{i\vec{k}^{(\alpha )}{\mathbf{r}}%
\}u_{\alpha }({\mathbf{r}})$\} are Bloch waves ($\alpha =1,...,n$) excited in
the target, $\ \{u_{\alpha }({\mathbf{r}})\}$ are periodical functions: $%
u_{\alpha }({\mathbf{r}}+{\mathbf{l}}_{m})=u_{\alpha }({\mathbf{r}})$, where ${\mathbf{l}}_{m}$
are the translation vector of the periodic structure, $k_{bz}=k_{z}\sqrt{1+%
\frac{\omega _{L}^{2}a_{0}^{2}}{\gamma ^{3}\Delta ^{2}D_{0}}}$, ${\mathbf{k}}%
_{b}=({\mathbf{k}}_{\perp };k_{bz})$, ${\mathbf{k}}_{b}^{(-)}=({\mathbf{k}}_{\perp
};-k_{bz})$ are the wave vector corresponding to an electromagnetic waves in
the electron beam. ${\mathbf{k}}_{b}$ and ${\mathbf{k}}_{b}^{(-)}$ are produced as the
solution of dispersion equation for electromagnetic waves in the beam, $%
\omega _{b}^{2}=4\pi n_{b}/m_{e}$ is the Langmuir frequency of \ electron
beam.

We assume that only the wave with wave vectors ${\mathbf{k}}$ and ${\mathbf{k}}^{(-)}$
are under the Cherenkov synhronism conditions with the particles. Therefore
the electron beam does not affect the diffracted waves with the wave vectors 
${\mathbf{k}}_{i}={\mathbf{k}}+{\mathbf{\tau}}_{i}$ if ${\mathbf{\tau}}_{i}\neq 0$. $a$, $b$, $%
\{m_{i}\}$, $c$, $d$, $t$, $\{f_{\alpha }\}$, $\{g_{i}\}$ are the
coefficients defined from boundary conditions on the surfaces of
discontinuity.Using equations (\ref{bound2})  
the following system for these coefficients can be written:

\begin{equation}
\begin{array}{c}
f=a\exp (ik_{bz}H)+b\exp (-ik_{bz}H)+\beta \frac{\omega _{L}^{2}}{\gamma }%
\frac{uk_{z}\eta }{\omega D_{0}}\left\{ a\exp (ik_{bz}H)-b\exp
(-ik_{bz}H)\right\}  \\
f=s\left\{ a\exp (ik_{bz}H)-b\exp (-ik_{bz}H)\right\}  \\
a\exp (ik_{bz}h)+b\exp (-ik_{bz}h)+\beta \frac{\omega _{L}^{2}}{\gamma }%
\frac{uk_{z}\eta }{\omega D_{0}}\left\{ a\exp (ik_{bz}h)-b\exp
(-ik_{bz}h)\right\} =\\
c\exp (ik_{z}h)+d\exp (-ik_{z}h)  \\
s\left\{ a\exp (ik_{bz}h)-b\exp (-ik_{bz}h)\right\} =c\exp (ik_{z}h)-d\exp
(-ik_{z}h)   \\
{\mathbf{...}}   \\
\text{where \ \ }s=\frac{k_{bz}}{%
k_{z}\left\{ 1+\frac{\omega _{L}^{2}a^{2}_0}{\gamma ^{3}\Delta ^{2}D_{0}}%
\right\} },\text{ \ \ \ }\eta =\frac{ck_{bz}}{\omega \left\{ 1+\frac{%
\omega _{L}^{2}a_{0}^{2}}{\gamma ^{3}\Delta ^{2}D_{0}}\right\} }  
\end{array} \label{linsys}
\end{equation}
The conditions on the beam boundaries are written in (\ref{linsys}), the
dots $...$ denote remaining boundary conditions on the surfaces of slow wave
system. Resolving (\ref{linsys}) it can be produced the following equality

\begin{eqnarray}
d &=&\exp (-2\alpha h)[1-s^{2}+2\eta _{1}]\frac{\exp (\alpha _{b}\delta
)-\exp (-\alpha _{b}\delta )}{(s+1)^{2}\exp (\alpha _{b}\delta
)-(s-1)^{2}\exp (-\alpha _{b}\delta )}a,  \label{connection} \\
\text{where }\eta _{1} &=&\beta \frac{\omega _{L}^{2}uck_{z}^{2}}{\gamma
\omega ^{2}D_{0}}  \nonumber
\end{eqnarray}
Here $\alpha = k_z/i,\alpha _{b}=k_{bz}/i$.
Let us note that roots of equation $d=0$ gives the eigenstate of ''cold''
waveguide without an electron beam. Therefore the generation equation for
the system ''electron beam + slow-wave system'' looks like

\begin{eqnarray}
-\exp (-2\alpha h)\left\{ \frac{\omega _{L}^{2}}{\gamma ^{3}\Delta ^{2}}+%
\frac{\omega _{L}^{2}}{\gamma ^{3}D_{0}}\right\} \frac{\exp (\alpha
_{b}\delta )-\exp (-\alpha _{b}\delta )}{(s+1)^{2}\exp (\alpha _{b}\delta
)-(s-1)^{2}\exp (-\alpha _{b}\delta )}=  \label{generation} \\
N({\mathbf{k}},{\mathbf{k}}_{1},...,{\mathbf{k}}_{n},\omega )  \nonumber
\end{eqnarray}
In (\ref{generation}) the function $N({\mathbf{k}},{\mathbf{k}}_{1},...,{\mathbf{k}}%
_{n},\omega )$ describes the ''cold'' slow-wave system. It is easy to see
distinction between lasing in cases with low and strong guiding field from (%
\ref{generation}). For strong guiding field (\ref{generation}) has form 
\begin{equation}
-\exp (-2\alpha h)\frac{\omega _{L}^{2}}{\gamma ^{3}\Delta ^{2}}\frac{\exp
(\alpha _{b}\delta )-\exp (-\alpha _{b}\delta )}{(s+1)^{2}\exp (\alpha
_{b}\delta )-(s-1)^{2}\exp (-\alpha _{b}\delta )}=N({\mathbf{k}},{\mathbf{k}}_{1},...,%
{\mathbf{k}}_{n},\omega )  \label{strong}
\end{equation}
and for the slow magnetic field 
\begin{equation}
-\exp (-2\alpha h)\frac{2\omega _{L}^{2}}{\gamma ^{3}\Delta ^{2}}\frac{%
\exp (\alpha _{b}\delta )-\exp (-\alpha _{b}\delta )}{(s+1)^{2}\exp (\alpha
_{b}\delta )-(s-1)^{2}\exp (-\alpha _{b}\delta )}=N({\mathbf{k}},{\mathbf{k}}_{1},...,%
{\mathbf{k}}_{n},\omega )  \label{slow}
\end{equation}
The terms related with electron beam differ in two times. It is result of
transverse dynamics lack in the case of (\ref{strong}). In the case of slow
guiding magnetic field (\ref{slow}) the transverse motion and longitudinal
motion give the same contribution to generation process.\bigskip

\section{Increments of quasi-Cherenkov instability}

\bigskip The slow electromagnetic wave produces the modulation in the
electron current and this modulation forms the coherent quasi-Cherenkov
radiation which acts on the electron beam again. As the result the emission
increases during the process of ''electron beam - radiation'' interaction in
the slow-wave system. Dynamics of this process can be described by the
increment of instability. Received generation equation (\ref{generation})
will be used for calculation of increment. 

Expanding the equation (\ref{generation}) in vicinity of Cherenkov resonance
and in vicinity of the eigenmode of   slow-wave system the generation equation can be
written in the form

\begin{equation}
-A\left\{ \frac{1}{\nu ^{2}}+\frac{1}{\nu ^{2}-a_{0}^{2}}\right\} =N_{0}+%
\frac{\partial N}{\partial \nu }\nu +\frac{\partial ^{2}N}{\partial \nu
^{2}}\nu ^{2}+...  \label{extrapolation}
\end{equation}
where $A=\exp (-2\alpha h)\frac{\omega _{L}^{2}}{\gamma ^{3}\omega ^{2}}$, $%
a_{0}=\frac{eH^{(0)}}{m\gamma c\omega }$. The dependence of increment on
the value of magnetic field can be studied using (\ref{extrapolation}). Let
discuss the physical meaning of the terms in right hand side of (\ref
{extrapolation}). As was shown in \cite{3}, the  $N_{0}$ is proportional to
absorption losses of slow-wave system ($\sim \chi _{0}^{\prime \prime }$). $%
\frac{\partial N}{\partial \nu }$ is equal to zero at the point of root
degeneration. In the case of great photoabsoption losses ($|N_{0}|\gg |%
\frac{\partial N}{\partial \nu }\nu |$) the dissipative instability 
developes.  At the root
degeneration point  dependence of increment on current density
changes $\nu \sim j^{1/(2+s)}$, where $s$ is the number of degenerated
modes.The dependence of increment on the magnitude of guiding field is
presented on (Figure \ref{fig.3}, Figure \ref{fig.4}) for current density $%
j=10$ $A/cm^{2}$ (Figure \ref{fig.3}) and $j=100$ $A/cm^{2}$\ (Figure \ref
{fig.4}). The following parameters were taken: $\omega \sim 5$ $\cdot
10^{11}s^{-1}$, $u\sim 1.4\cdot 10^{10}cm/s$. It follows from (Figure \ref
{fig.3}, Figure \ref{fig.4}) that for the magnitude of magnetic $H_0=3$ KGs
only longitudinal motions of electron contributes to stimulated emission if
the current density  $j=10$ $A/cm^{2}$. If  $j=100$ $A/cm^{2}$, the critical
magnitude of the guiding field is $H_0\approx 6$ KGs. If guiding field less
these values, the transverse dynamics contributes to emission also.

\section{Conclusions}

\begin{figure}[htbp]
\epsfxsize=15 cm
\epsfbox{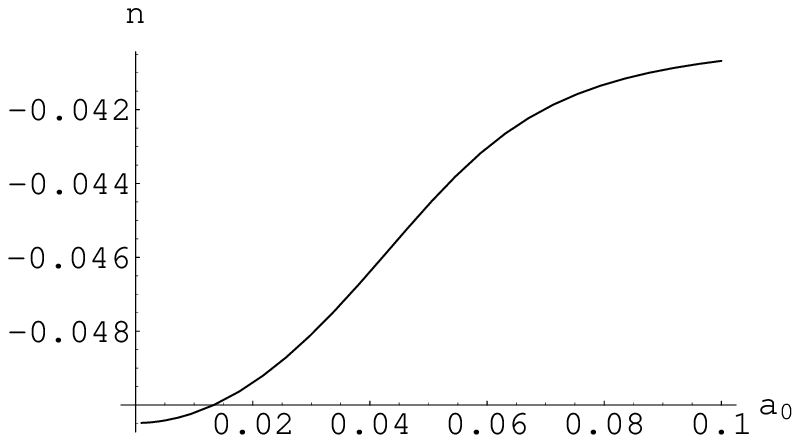}
\caption{
Dependence of dimensionless instability increment $\nu$ on the magnitude of guiding field 
($a_0 =\frac{e H_0}{m c\gamma \omega} $). The current density of electron beam is $j=10~A/cm^2$. }
\label{fig.3}
\end{figure}

\begin{figure}[htbp]
\epsfxsize=15 cm
\epsfbox{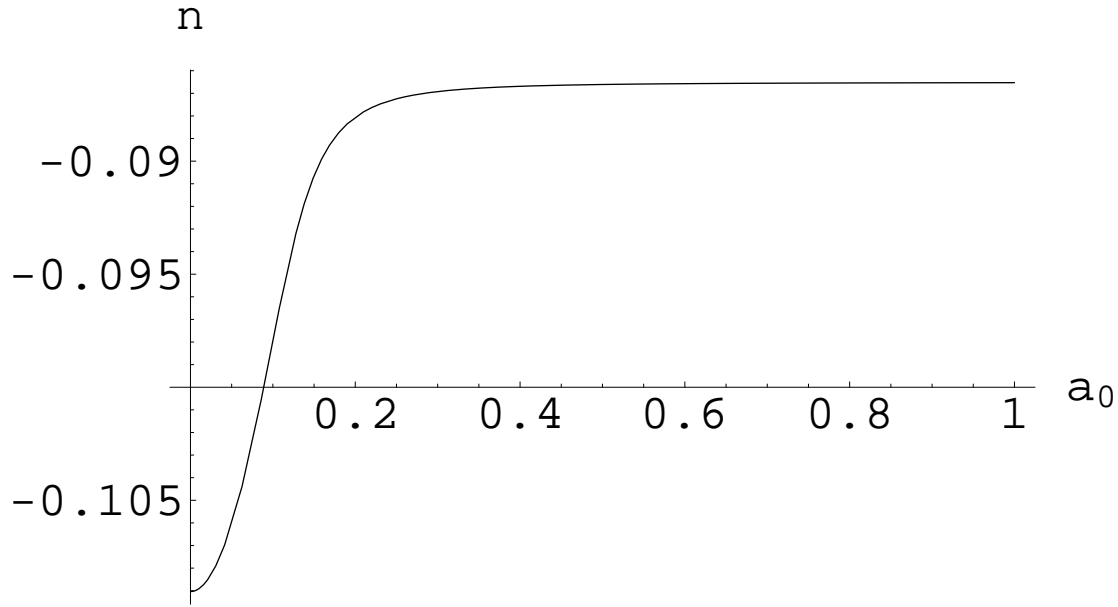}
\caption{
Dependence of dimensionless instability increment $\nu$ on the magnitude of guiding field 
($a_0 =\frac{e H_0}{m c\gamma \omega} $). The current density of electron beam is $j=100~A/cm^2$.
 }
\label{fig.4}
\end{figure}

This paper presents analysis of the guiding magnetic field influence
 on the quasi-Cherenkov stimulated radiation. It is shown that the
increment is maximal in the case without magnetic field. However, for
guiding of the electron beam over the surface of the slow-wave system the
magnetic field has to be strong enough to oppose to Coulomb repulsion of
electron beam.  The following simple estimation for guiding
field can be used: 1) the deviation from the Cherenkov  synchronism $\Delta $
must be less the width of stimulated emission line 
\begin{equation}
|\delta \Delta |\sim \omega \delta v_{x}/u+\omega \delta v_{z}/\gamma u\leq
\max \{\pi u/L_{int};\beta \omega \nu \},  \label{spread}
\end{equation}
where $\delta v_{x}$ and $\delta v_{z}$ are velocity pertubations caused by
Coulomb repulsion, 2) the amplitude of oscillation in crossed fields can
be less then the gap width $h $: $\frac{mc^{2}E}{eH^{2}}<h $.
If period of transverse electron oscillations in guiding magnetic field less then interaction time 
($\omega _{H}L/v\geq 1$), then (\ref{spread}) can be writen as $\omega
cE/\gamma ^{2}uH\leq \max \{\pi u/L_{int};\beta \omega \nu \}$. Take
into account that $E\sim \frac{2\pi I}{ul}$, $2\pi \omega cI/(u^{2}lH)\leq
\max \{\pi u/L_{int};\beta \omega \nu \}$, $\frac{2\pi mc^{2}I}{%
eH^{2}ul\gamma ^{2}}<h $ where $l$ is the width of an electron beam
along the $y$ axis, $I$ is the beam current.So, if the inequalities max\{$%
\frac{eH^{(0)}}{m\gamma c};2\pi \omega cI/(\gamma ^{2}u^{2}lH)\}<\max \{\pi
u/L_{int};\beta \omega \nu \}$ are fulfilled, then the transverse dynamics
of electrons contributes to generation process. As was shown above the
transverse dynamics contributes to stimulated emission for magnitude of
guiding magnetic field $\leq $ few $KGs$. The transverse electron dynamics
can increase the increment on 25 percent in the high gain exponentional
regime as can be seen from (Figure \ref{fig.3}, Figure \ref{fig.4}). In the
slow gain regime the magnitude of the gain can increase in two times due to
transverse dynamics.

\begin{figure}[htbp]
\epsfxsize=15 cm
\epsfbox{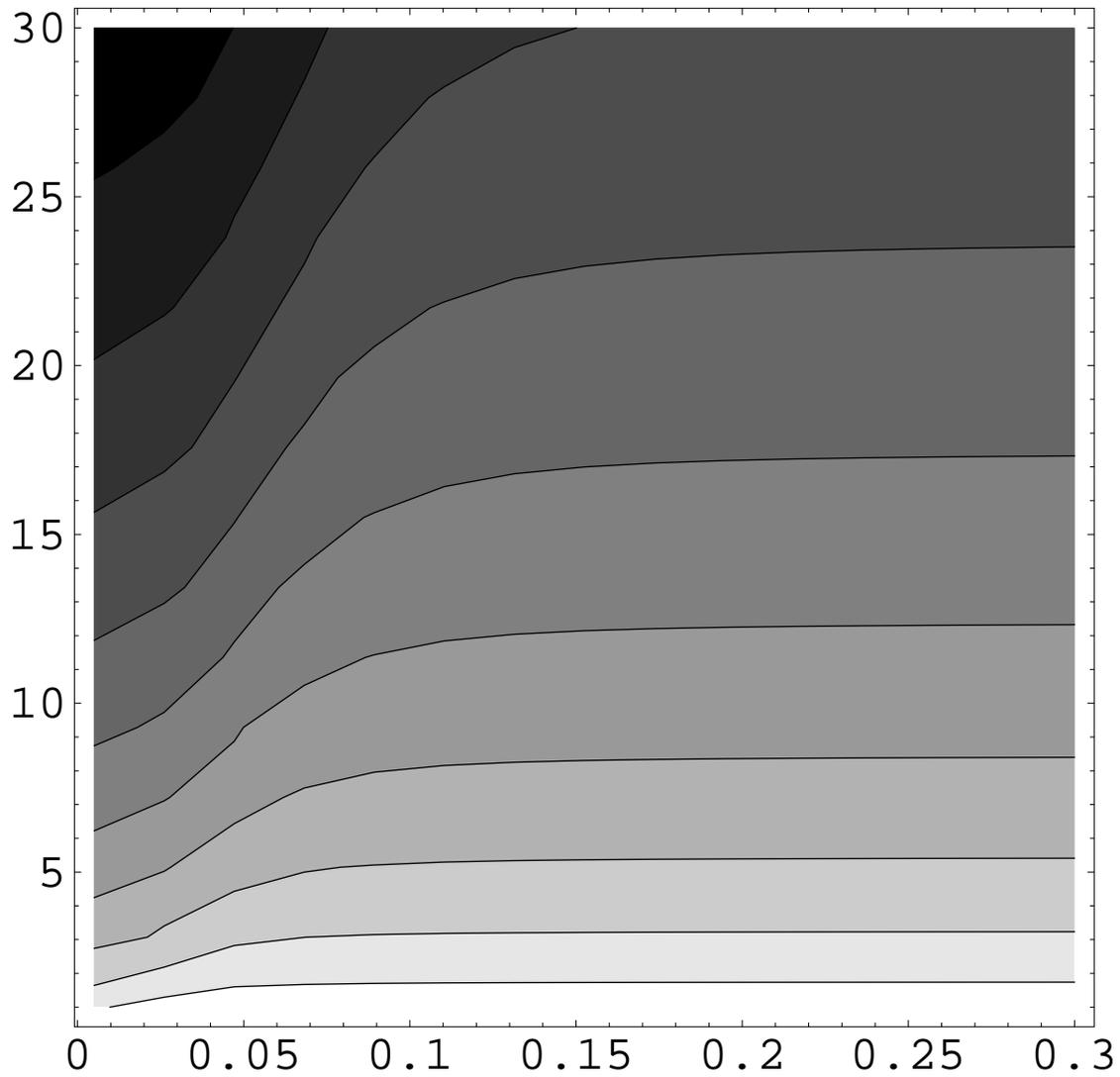}
\caption{
The contour plot for instability increment.The abscissa corresponds to $a_0$, the ordinate corresponds to current density $j~(A/cm^2) $.
 }
\label{fig.5}
\end{figure}
\end{document}